\documentstyle[preprint,aps]{revtex}
\begin{document}
\preprint{ SUNY-RHI-94-11, subm. to Phys. Lett. B}
\normalsize
\title { Thermal Equilibration and Expansion in Nucleus-Nucleus Collisions
   at the AGS}
\author{P. Braun-Munzinger, J. Stachel, J. P. Wessels and N. Xu$^*$}
\address {\em Department of Physics \\
       State University of New York at Stony Brook \\
       Stony Brook, \, New York 11794 -- 3800 }
\date{\today}
\maketitle
\begin{abstract}
The rather complete data set of hadron yields from central Si + A
collisions at the Brookhaven AGS is used to test whether the system at
freeze-out is in thermal and hadro-chemical equilibrium. Rapidity and
transverse momentum distributions are discussed with regards to the
information they provide on hydrodynamic flow.
\end{abstract}
\narrowtext

The goal of the ultra-relativistic heavy-ion program at the BNL AGS and
CERN SPS is to study highly excited and dense nuclear matter and
possibly the transition from hot and dense hadronic matter to deconfined
quark-matter with restored chiral symmetry. While future collider experiments
will probe a hot quark-gluon plasma with low net baryon
density, present fixed target experiments create matter, possibly
quark-matter, at very high baryon density and moderate temperature.

The present paper is following up on an earlier suggestion by some of us
\cite{jspbm}, based on the first AGS and SPS data, that a high degree
of thermalization is reached and that there is evidence for hydrodynamic
expansion of the created fireball.  We now  use the much larger
set of data from central Si + A collisions at the AGS first to discuss
quantitatively the validity of a thermodynamic approach to interpret the
data. Within this approach present data allow to determine the
temperature of the system at the point when particles seize to interact
strongly (freeze-out) as well as, via the net baryon density at
freeze-out, the baryon chemical potential. The relatively large
freeze-out temperature thus determined implies that even higher
temperatures are reached earlier in the collision. In order to reach the
freeze-out stage, the system has to expand
considerably. Longitudinal and transverse spectra and, in particular,
their mass dependence can yield information on the expansion
velocity. This discussion forms the second part of this paper.

The present study starts on the following background:\\
i) Production of transverse energy and the proton rapidity
distribution after a central Silicon nucleus collision indicate a
high degree of stopping \cite{E814-et,E814-1,802prc,810p,jspbm,rqmd1}.\\
ii) Hadronic cascade codes that reproduce the hadron distributions at
freeze-out require baryon densities in excess of five times
normal nuclear matter density for an extended time,
typically about 5 fm/c \cite{rqmd1,arc1}.\\
iii) Data on pion interferometry are consistent with the fireball
created in Si + Pb central collisions having a large final transverse
radius of 6.7 fm (2.7 times the Si transverse radius), and a duration of
the pion emission of 9 fm/c \cite{E814_hbt}.

For this fireball scenario we explore quantitatively the predictions of
a consistent thermodynamic and hydrodynamic approach to describe the AGS
data. Calculations using ideal gas thermodynamics have been reported
before \cite{cley_ags,cley_cern,raf_ags} and compared to particle yield
ratios. We use basically the same formalism as
\cite{cley_ags,cley_cern}. However, the authors of \cite{cley_ags} had a
much smaller set of early AGS and SPS data to compare to and our
philosophy in fixing the model parameters is different. In contrast to
\cite{raf_ags}, we use chemical equilibrium throughout combined with
strangeness conservation.

To describe particle distributions and abundance at freeze-out, we need
to treat the thermodynamics of the system only after the initially hot
and dense fireball has expanded and reached the freeze-out density. In
every co-moving restframe the system is therefore described by a grand
canonical ensemble of non-interacting fermions and bosons in equilibrium
at freeze-out temperature $T$. For an infinite volume the particle
number densities are given as integrals over particle momentum $p$:
\begin{equation}
 \rho_i^0 = \frac{g_i}{2\pi^2} \int_{0}^{\infty} \frac{p^2 dp}{{\rm
exp}[(E_i-\mu_b B_i - \mu_s S_i)/T] \pm 1}
\end{equation}
where $g_i$ is the spin-isospin degeneracy of particle {\it i}, $E_i$,
$B_i$ and $S_i$ are its total energy in the local restframe, baryon
number and strangeness, and $\mu_b$ and $\mu_s$ are the baryon and
strangeness chemical potentials ($\hbar$=c=1 unless otherwise
noted). For a system of finite size the argument of the integral in
equation (1) has to be multiplied \cite{finvol} by a correction factor
which we evaluate for a spherical volume with radius $R$. We also apply
the excluded volume correction \cite{cley_cern} to take into account the
volume occupied by individual baryons and mesons with radii of 0.8 and
0.6 fm.

The temperature range relevant for the present study is 0.10 - 0.15 GeV
which sets the scale for the mass range of particles to be
considered. We include strange and non-strange mesons up to a mass of 1.5
GeV and baryons up to 2 GeV. Results change by less than 10\% if the
mass range is restricted to mesons lighter than 1 GeV and baryons
lighter than 1.5 GeV, indicating the sensitivity to the mass cut. We
have omitted hyperons with strangeness 2
or larger {\it and}
mass above 1.6 GeV since their yields and decays will not impact on any
of the presently observed quantities. It is very important, however, to
treat particles and antiparticles as well as different isospin states
evenly, i.e. to include all states at a given energy, since observables
like the $\bar{\Lambda}/\Lambda$ or ${\rm \bar{p}/p}$ ratio are strongly
affected by feeding from higher states.

The range of temperatures to be considered is driven by the experimental
observation \cite{814pi} of the occupation probability of the
$\Delta(1232)$ resonance in E814: For a system in thermal equilibrium,
temperatures of 0.12 - 0.14 GeV are consistent with the observed
abundance. The baryon chemical potential is constrained by the measured
pion to nucleon abundance as well as the density of the system at
freeze-out \cite{E814_hbt}. We choose a value of
$\mu_b$ = 0.54 GeV. For a given temperature and baryon chemical
potential the strangeness chemical potential is fixed by strangeness
conservation. In particular, for $T$ = 0.120 and 0.140 GeV one obtains
$\mu_s$ = 0.108 and 0.135 GeV. Using these input parameters and
equation (1) we find primary particle densities.

For comparison to experimental data one
needs to consider decay and feeding. We use all the known
branching ratios as given in \cite{databook} as well as symmetry and phase
space arguments for unknown branching ratios. To illustrate the importance
of treating the feeding properly we note that, at $T$ = 0.14 GeV, the
primary pion yield is tripled and the nucleon yield is doubled by
feeding from higher resonances.

Flow effects discussed below do not affect angle integrated particle
densities, but may severely change particle densities and density ratios
at fixed rapidity. We, therefore, compare predictions of the thermal
model to experimental quantities integrated over transverse momentum
$p_t$ and rapidity $y$. The $y$ integration is justified since even a
rather well localized thermal source in the center of the colliding
nuclei is spread out in $y$ because of the width of the space-rapidity
correlation and the natural width in $y$ of a thermal source. Also the
data cover at most 2 units of $y$ and all include midrapidity.

In Table\ \ref{ratios} all currently available experimental data on
particle ratios measured in central Si + Au(Pb) collisions are compared
to the corresponding ratios calculated for two temperatures, 0.12 and
0.14 GeV. Overall good agreement is found. Although the yields vary
over three orders of magnitude, all particle ratios are reproduced to
better than a factor of two. While our choice for the temperature range
considered is driven by the $\Delta(1232)$ abundance and not the yield
ratios in Table\ \ref{ratios}, the range considered nevertheless gives
the best overall agreement; the experimental d/(p+n), K/$\pi$,
$\Lambda$/(p+n) and $\phi$/K ratios favor a slightly lower temperature,
the \=p/p and $\bar{\Lambda}/\Lambda$ ratios are bracketed by the range
considered, the K$^+$/K$^-$ and $\Xi^- / \Lambda$ ratios favor higher
$T$.

The choice of $\mu_b$ is a trade-off between the observed pion to
nucleon ratio and the particle density at freeze-out. At a given
$\mu_b$, absolute particle densities predicted by the model are very
sensitive to the exact freeze-out temperature; for temperatures of $T$ =
0.12 (0.14) GeV and $\mu_b$ = 0.54 GeV, densities for nucleons and pions
of $\rho_n = 0.070~(0.13) /{\rm fm^3}$ and $\rho_{\pi}= 0.09~(0.17)
/{\rm fm^3}$ are obtained. These numbers should be compared with
$\rho_n^{exp} = 0.058/{\rm fm^3}$ and $\rho_{\pi}^{exp} = 0.063/{\rm
fm^3}$ recently estimated \cite{E814_hbt} for Si + Pb central
collisions. Especially for the lower temperature the agreement for the
absolute densities is surprisingly good. From the data displayed in
Table\ \ref{ratios} there is no indication that strangeness is not in
chemical equilibrium.

In the following we will discuss whether the experimental rapidity and
transverse momentum spectra are consistent with the prediction of such a
thermodynamic model allowing for possible flow effects. Equation (1)
implies random, isotropic emission. Since the temperature at freeze-out
exceeds 100 MeV, we use the Boltzmann approximation. Transformed into
rapidity $y$ and transverse momentum $p_t$ this implies:
\begin{equation}
 E \frac{d^3N}{d^3p} \propto E \exp(-E/T) = m_t {\rm cosh}(y) \exp(-m_t
{\rm cosh}(y)/T)
\end{equation}
for a particle with mass $m$ and transverse mass $m_t$. All kinematical
variables are evaluated in the nucleon-nucleon center of momentum frame.
Integrating over $m_t$, one obtains:
\begin{equation}
\frac{dN_{iso}}{dy} \propto m^2 T(1 + 2\chi + 2\chi^2) \exp(-1/\chi)
\end{equation}
with $\chi = T/(m {\rm cosh}(y))$. For pions with $m \approx T$, this
distribution is close to that for massless particles, {\it i.e.}
proportional to $1/{\rm cosh}^2(y)$. For heavier particles isotropic
emission implies a strong narrowing of the distribution, in
contradistinction to experimental observations.  This is demonstrated in
Fig.\ \ref{fig1}, where experimental rapidity distributions
\cite{E814-1,802prc,810plb,814pi1} for central collisions of Si + Al are
compared to predictions of the isotropic thermal model (solid
line). This rather small system was chosen since it is the only
symmetric system where data for particle distributions are
published. The calculations were performed for $T$ = 0.12 GeV; the
situation is very similar at 0.14 GeV.  Obviously the agreement between
the isotropic model calculation and data is poor. Much better agreement
between data and calculations is obtained by introducing a common
collective flow velocity in longitudinal direction. Following
\cite{sashar,ekkard} we superpose individual isotropic thermal
sources within a rapidity interval [$-y'_{max},y'_{max}$]:
\begin{equation}
\frac{dN}{dy} =
\int_{-y'_{max}}^{y'_{max}} dy' \frac{dN_{iso}(y-y')}{dy'}.
\end{equation}
The integration limit $y'_{max}$ confines boost invariance to a finite
rapidity interval; we treat it as a free parameter to determine the
amount of flow required by the data. The results of calculations for
such a longitudinally expanding fireball are shown as dashed lines in
Fig.\ \ref{fig1}. Good overall agreement with the data is found for
$y'_{max}=1.15$. Averaged over the forward (backward) portion of the
flat distribution (4) this corresponds to $\langle y' \rangle = 0.58$
(-0.58) and a mean longitudinal flow velocity of $\langle \beta_l
\rangle = {\rm tanh}(\langle y' \rangle) = 0.52$. We note that, although
the normalization of the calculated curves has been adjusted for each
particle species separately, the normalization factors in principle
could have been taken from our calculations above. This would imply,
{\it e.g.} an upward shift of 20\%, 37\% and 60\% for p, $\pi^+$, and
K$^+$. Considering the uncertainty in the freeze-out volume, this
agreement is remarkably good.

Rapidity distributions of d, $\Lambda$, p, K are not strongly affected
by resonance decays \cite{ekkard}. The situation may be different,
however, for pions: as demonstrated elsewhere \cite{814pi}, at AGS
energies about 1/3 of all pions originate from decay of the
$\Delta(1232)$ resonance. This will also widen somewhat the pion
rapidity distributions, and may account for part of the effect flow has
on the pion distributions.

The width of the rapidity distribution for protons may also be an
indication for incomplete stopping in this rather small system
\cite{E814-1}. From the difference in the p and
$\Lambda$ distributions one can infer the degree of transparency. The
$\Lambda$ distribution in Fig.\ \ref{fig1} does not support $\langle
\beta_l \rangle$ larger than 0.52 indicating that the
protons are slowed down 0.2 units of rapidity less than required for
full stopping. Overall, thermalization plus longitudinal flow
provide a fairly consistent description of all rapidity distributions
already for the relatively small system Si + Al.

Assuming that the longitudinal and transverse motion of the thermal
source are decoupled, flow effects can be incorporated into transverse
momentum spectra following \cite{ekkard}:
\begin{equation}
\frac{dN}{m_tdm_t} \propto
\int_{0}^{R} r dr \, m_t I_0(\frac{p_t{\rm sinh}(\varrho)}{T})
K_1(\frac{m_t{\rm cosh}(\varrho)}{T})
\end{equation}
with Bessel functions $I_0$ and $K_1$, a parameter $\varrho = {\rm
tanh}^{-1}(\beta_t)$ and a transverse velocity profile of $\beta_t(r) =
\beta_t^{max}(r/R)^{\alpha}$.  The flow parameter $\beta_t^{max}$
determined from data for ($m_t-m) > 0.3\,{\rm GeV/c^2}$ is not very
sensitive to the magnitude of the parameter $\alpha$; we have used
$\alpha = 1$ in the subsequent calculations. Since we will focus our
analysis on Si + Au data, we choose for the transverse freeze-out radius
of the system the value $R=6.7$ fm \cite{E814_hbt}.

Rather than exploring the parameter space ($T$, $\beta_t^{max}$)
describing the data, we fix the freeze-out temperature between 0.12 and
0.14 GeV. We then determine whether the data can be consistently
described with one common transverse expansion velocity by comparing
results of calculations using eq. (5) with data \cite{802prc} near
midrapidity.

To determine
best fit values for $\beta_t^{max}$, we restrict the fit to
$(m_t-m) > 0.3$ GeV/c$^2$. In this range resonance decays yield negligible
changes in spectra for d, p, and K. For pions  inclusion of resonance
decays makes a significant change  at
low $(m_t-m)$ values. We have, therefore, included resonance decays into
the calculation of transverse momentum spectra by numerical simulation
of two-body and three-body decays of the dominant resonances.

The results are presented in Fig.\ \ref{fig2}. Remarkable agreement
between data and calculations is obtained for $\beta_t^{max}$ = 0.58
(0.50) at $T$ = 0.12 (0.14) GeV, even for pion spectra at low
$m_t$. This corresponds to average flow
velocities of $\langle \beta_t \rangle$ = 0.39 (0.33) and implies
expansion times of the order of 12 fm/c.  Independent support for the
presence of flow effects was recently obtained \cite{flow_prl} in an
analysis of azimuthal distributions of transverse energy in semi-central
Au+Au collisions at AGS energies.

We have demonstrated that the presently available AGS data can be
consistently described in a thermal model with chemical equilibrium and
flow. This includes particle densities, ratios of produced particles,
and rapidity and transverse momentum distributions. The thermal
parameters describing freeze-out are $T$ = 0.12 - 0.14 GeV, $\mu_b=0.54$
GeV, $\langle \beta_l \rangle = 0.52$, and $\langle \beta_t \rangle =
0.39 - 0.33$. Earlier times in the evolution of the fireball need to be
probed with different observables to determine the equation of state
during the high density phase.

This work was supported by the NSF. One of us (J.P.W.) is supported by
the A. v. Humboldt Foundation as a F. Lynen fellow.

\noindent $^*$ present address: P-25, MS D456, LANL, Los Alamos, NM 87545.

\begin{table}
\caption{Particle ratios calculated in a thermal model for two different
temperatures, baryon chemical potential $\mu_b$= 0.54 GeV and
strangeness chemical potential $\mu_s$ such that overall strangeness is
conserved, in comparison to experimental data (with statistical errors
in parentheses) for central collisions of 14.6 A GeV/c Si + Au(Pb).
\label{ratios}}
\begin{tabular}{||c|ll|llc||}
Particles & \multicolumn{2}{c|} {Thermal Model} & \multicolumn{3}{c||}{Data}\\
 & $T$=.120 GeV & $T$=.140 GeV & exp. ratio & rapidity & ref.\\ \hline
$\pi$/(p+n) & 1.29 & 1.34 & 1.05(5) & 0.6 - 2.8 & \cite{802prc,E814-1}\\
d/(p+n)     & 4.3 $\cdot$ 10$^{-2}$ & 5.8 $\cdot$ 10$^{-2}$ & 3.0(3) $\cdot$
10$^{-2}$ & 0.4 - 1.6 & \cite{802prc} \\
\=p/p   & 1.47 $\cdot$ 10$^{-4}$ & 5.8 $\cdot$ 10$^{-4}$ & 4.5(5)
$\cdot$ 10$^{-4}$ & 0.8 - 2.2 & \cite{802qm93}\\ \hline
K$^+$/$\pi^+$ & 0.23 & 0.27 & 0.19(2) & 0.6 - 2.2 & \cite{802prc}\\
K$^-$/$\pi^-$ & 5.0 $\cdot$ 10$^{-2}$ & 6.2 $\cdot$ 10$^{-2}$ & 3.5(5)
$\cdot$ 10$^{-2}$ & 0.6 - 2.3 & \cite{802prc}\\
K$^0_s$/$\pi^+$ & 0.14 & 0.16 & 9.7(15) $\cdot$ 10$^{-2}$ & 2.0 - 3.5 &
\cite{810plb,802prc,814pi1}\\
K$^+$/K$^-$ & 4.6 & 4.3 & 4.4(4) & 0.7 - 2.3 & \cite{802prc}\\ \hline
$\Lambda$/(p+n) & 9.5 $\cdot$ 10$^{-2}$ & 0.11 & 8.0(16) $\cdot$ 10$^{-2}$
& 1.4 - 2.9 & \cite{810plb,802prc,E814-1} \\
$\bar{\Lambda}/\Lambda$ & 8.8 $\cdot$ 10$^{-4}$ & 3.7 $\cdot$ 10$^{-3}$
& 2.0(8) $\cdot$ 10$^{-3}$ & 1.2 - 1.7 & \cite{802qm93}\\ \hline
$\phi$/(K$^+$+K$^-$) & 2.4 $\cdot$ 10$^{-2}$ & 3.6 $\cdot$ 10$^{-2}$ &
1.34(36) $\cdot$ 10$^{-2}$ & 1.2 - 2.0 & \cite{802qm93}\\  \hline
 $\Xi^- / \Lambda$ & 6.4 $\cdot$ 10$^{-2}$ & 7.2 $\cdot$ 10$^{-2}$ &
0.12(2) & 1.4 - 2.9 & \cite{810casc}\\ \hline
 \=d/\=p & 1.1 $\cdot$ 10$^{-5}$ & 4.7 $\cdot$ 10$^{-5}$ &
1.0(5)$\cdot$ 10$^{-5}$ & 2.0 & \cite{858dbar}\\
\end{tabular}
 \end{table}

\begin{figure}
\caption{Rapidity distributions for central 14.6 A GeV/c Si+Al
collisions
[3,4,16,21]
 in comparison to
isotropic thermal distributions at $T$ = 0.12 GeV (solid lines) and
distributions for a source at the same temperature expanding with
$\langle \beta_l \rangle$ = 0.52 (dashed lines).}
\label{fig1}
\end{figure}

\begin{figure}
\caption{Experimental particle spectra
[4]
at $y$ = 1.3 compared to calculated spectra for a source at $T$ =
0.12 GeV expanding transversely with $\langle \beta_t \rangle$ =
0.39 (left) and a source at $T$ = 0.14 GeV and $\langle \beta_t
\rangle$ = 0.33 (right). The arrows indicate the beginning of the fit
region. For details, including the treatment of resonance decays, see text.}
\label{fig2}
\end{figure}

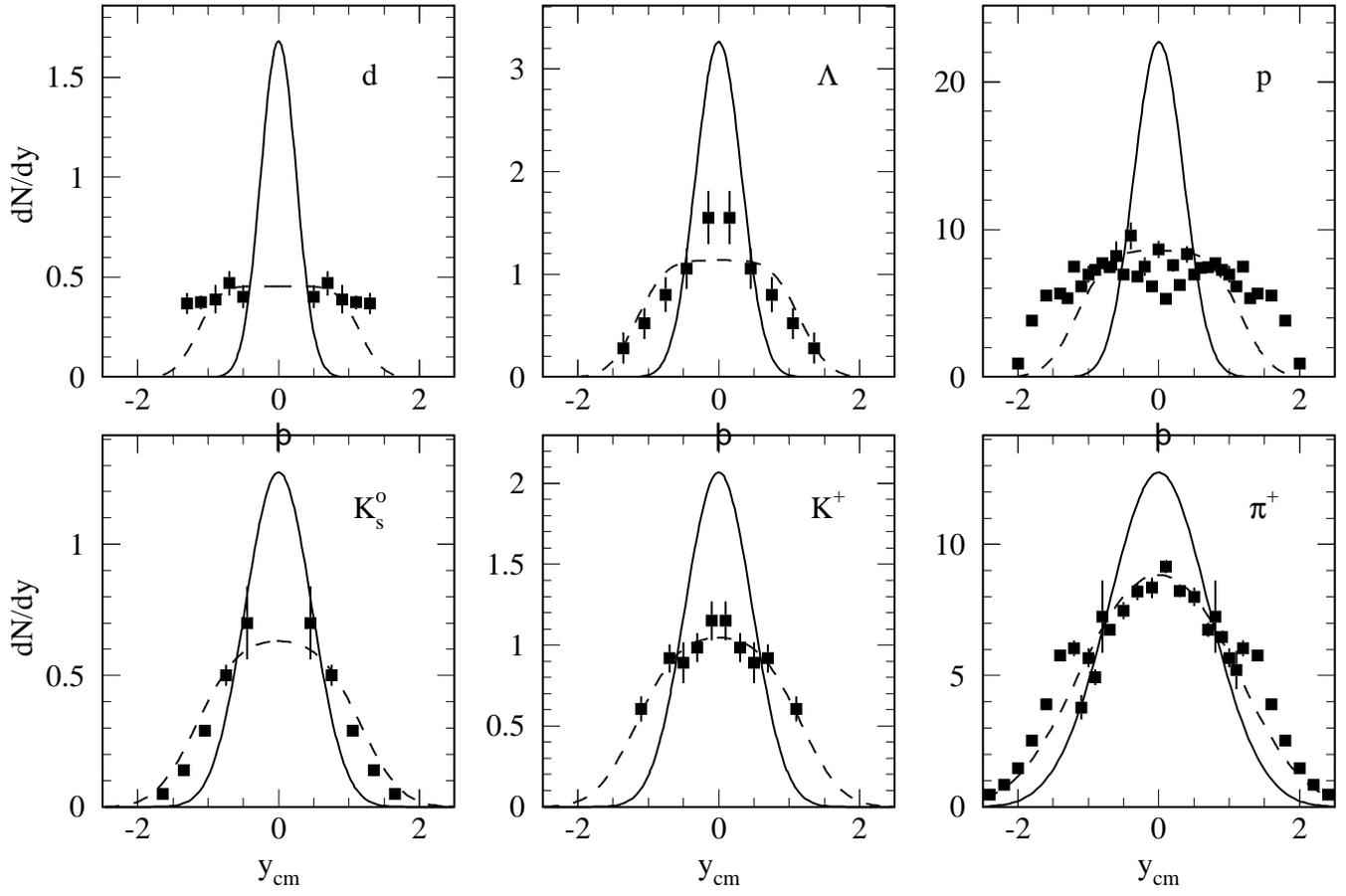
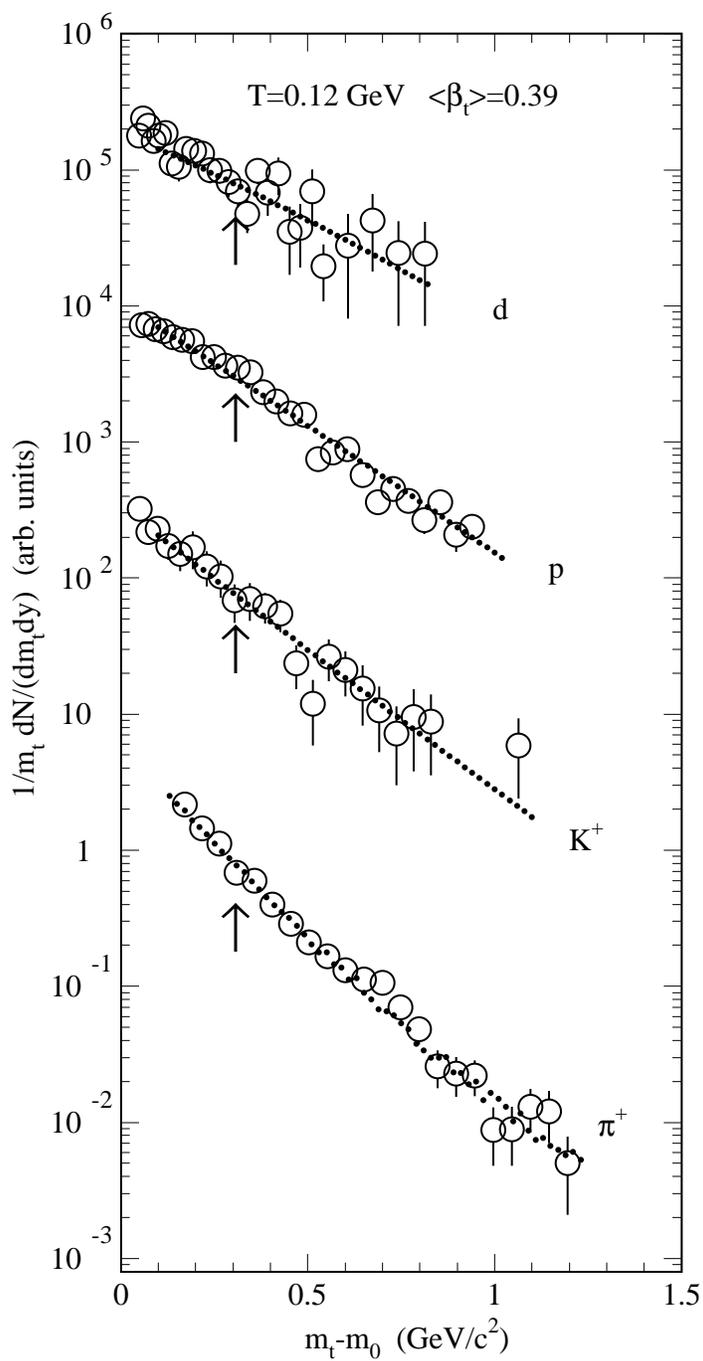
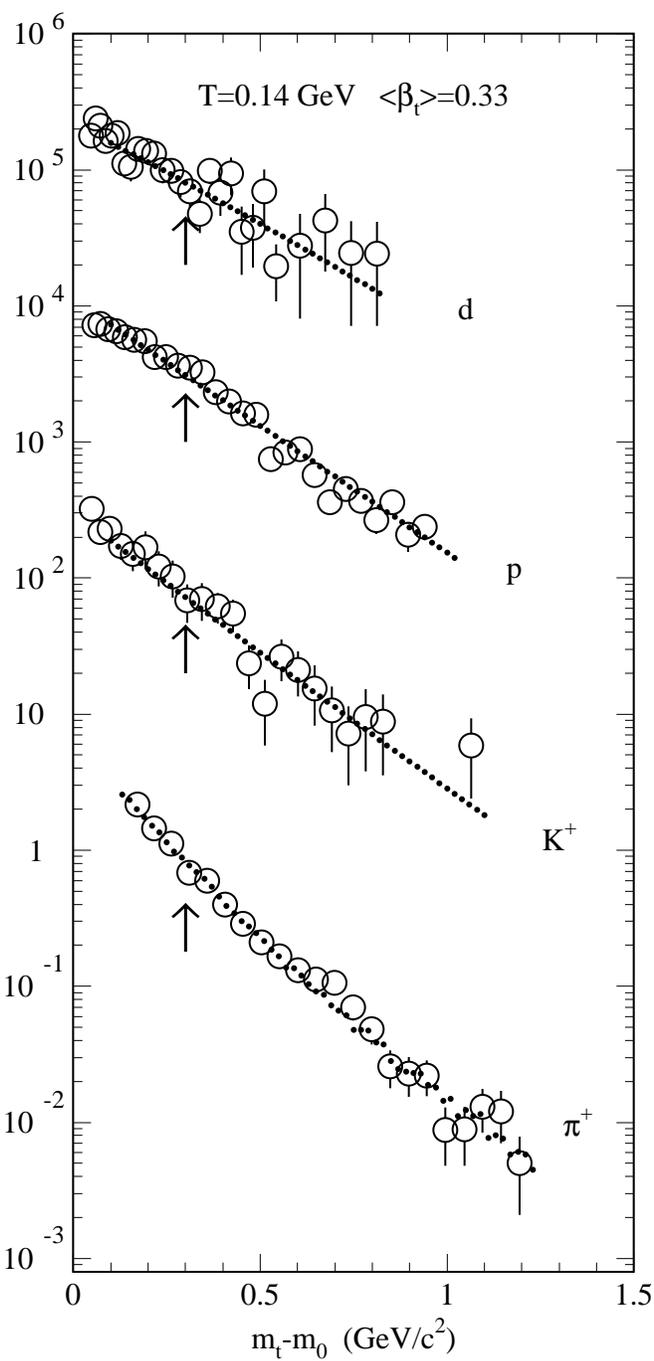
\end{document}